\documentclass[11pt]{article}

\usepackage[margin=1in]{geometry}
\usepackage{graphicx}
\usepackage{amsmath,amssymb,amsfonts}
\usepackage{mathrsfs}
\usepackage{xcolor}
\usepackage{tikz}
\usetikzlibrary{arrows.meta,positioning}
\usepackage{cite}
\usepackage[hidelinks]{hyperref}
\usepackage{orcidlink}

\newcommand{\savg}[1]{\left\langle #1 \right\rangle_{\Sigma}}

\title{Antipodal constraints on transient Hawking radiation in a conformal channel}
\author{%
M. Baran \"Okten$^{1,\dagger,*}$, Asl{\i} S. Turan$^{2,3}$\\[0.45em]
\small $^{1}$Faculty of Engineering and Natural Sciences, Sabanc{\i} University, Istanbul, T\"urkiye\\
\small $^{2}$Department of Physics, Bo\u{g}azi\c{c}i University, Istanbul, T\"urkiye\\
\small $^{3}$Department of Physics, Y{\i}ld{\i}z Technical University, Istanbul, T\"urkiye\\[0.2em]
\small $^{\dagger}$Present address: Department of Physics Engineering, Istanbul Technical University, Istanbul, T\"urkiye\\
\small $^{*}$Corresponding author: \href{mailto:okten26@itu.edu.tr}{okten26@itu.edu.tr}
}
\date{}

\begin{document}

\maketitle

\begin{abstract}
Hawking radiation away from stationarity is governed, at the ray tracing level, by the retarded time evolution of the null mapping. We consider an extracted massless conformal channel whose outgoing directions are labelled on a fixed reference horizon slice. Balanced uniformisation, after fixing a nondegenerate balanced branch, defines an induced antipodal involution on this slice. We prove that creation and erasure of centred antipodal odd structure in the channel labelled peeling field enforce a lower bound on the time integrated spatial norm of the conformal channel defect. The forcing scale depends only on the largest intermediate mismatch and the intrinsic Jacobian distortion of the induced antipode, while greybody filtering, angular and frequency mixing, mass effects, spin dependence, and residual propagation or extraction corrections enter through a remainder whose integrated norm must be controlled independently. Round sphere odd scalar harmonics provide analytic calibrations, with the compact dipole pulse saturating the geometric and temporal estimates. A time-averaged form identifies a transient conformal channel scale proportional to the antipodal peeling amplitude divided by the episode duration.
\end{abstract}


\section{Introduction}\label{sec:intro}

The equilibrium Hawking spectrum of a stationary black hole is commonly summarised by a single temperature set by the horizon surface gravity \cite{Hawking1975,Unruh1976,BirrellDavies1982,Wald1994,ParkerToms2009}. That description captures the stationary limit. In a generic nonstationary spacetime, particle creation follows the comparison between ``in'' and ``out'' mode decompositions, with the relevant information encoded in the global redshift relation between null parameters \cite{BroutMassarParentaniSpindel1995,Visser2003,BarceloLiberatiSonegoVisser2011}. In solvable \(1+1\) models and in moving-mirror analogues, this dependence on the null mapping becomes especially explicit, since the same ray-tracing data directly control the renormalised flux and the time dependence of particle production \cite{RST1992,CarlitzWilley1987,GoodAndersonEvans2013,GoodAndersonEvans2016,GoodYelshibekovOng2017}.

A natural transient episode motivating the construction is asymmetric perturbation of a black hole by infall or tidal forcing. In such situations the direction resolved peeling field $\kappa(u,\hat n)$ may acquire a transient odd component associated with the preferred direction introduced by the perturbation, in analogy with the preferred-axis horizon distortions seen in perturbed and tidally deformed black-hole settings \cite{Hartle1973,Hartle1974,FangLovelace2005,VegaPoissonMassey2011,OSullivanHughes2014,Poisson2015}. The centred antipodal mismatch introduced below converts the largest intermediate odd structure, together with its disappearance at the endpoints, into a profile-independent lower bound on the time-integrated conformal channel norm. The present analysis uses the asymptotic null ray tracing relation, or equivalently its peeling field, as the channel data controlling the conformal flux. Within a controlled effectively \(1+1\) conformal channel, the renormalised outgoing flux is an exact functional of that ray-tracing map, and the departure from the quadratic reference term is fixed by the retarded-time variation of the peeling field \cite{FullingDavies1976,DaviesFulling1976,CarlitzWilley1987,ChristensenFulling1977,FabbriNavarro2005,DiFrancesco1997,GoodAndersonEvans2016}. Retarded time variation of the peeling field generates the derivative contribution to the conformal flux. Closely related moving-mirror models display the same mechanism in a particularly transparent form, while also exhibiting transient positive-energy flares, negative-energy flux, finite-energy evaporation, and nontrivial entropy evolution \cite{GoodOng2015,GoodLinder2018,GoodLinder2019,BianchiSmerlak2014,BianchiSmerlak2014b,BianchiDeLorenzoSmerlak2015}.

A fixed reference horizon slice \(\Sigma\) provides the intrinsic geometry and channel label space on which antipodal structure is defined. The same abstract copy of this slice is used throughout the transient episode, and its points \(x\in\Sigma\) label outgoing channels through a physical extraction map \(\Phi_u\). From the direction resolved ray tracing map \(p(u,\hat n)\), the extraction map defines the channel labelled map \(p_{\Sigma}(u,x)=p(u,\Phi_u(x))\) and its peeling field \(\kappa_{\Sigma}(u,x)=-\partial_u\ln(\partial_u p_{\Sigma}(u,x))\), where the derivative is taken at fixed channel label \(x\). Balanced uniformisation defines an antipodal comparison map from the intrinsic metric $g$, allowing a centred mismatch functional that remains meaningful away from spherical symmetry \cite{Jost2006,OsgoodPhillipsSarnak1988,Hersch1970,DouadyEarle1986}. All norms, gradients, Jacobians, and spectral quantities use the intrinsic metric \(g\) on \(\Sigma\), while the round sphere supplies the comparison antipode. Spectral geometry then converts the mismatch into an explicit angular-gradient cost, while the temporal bound retains the intrinsic Jacobian distortion of the induced antipode \cite{Chavel1984,Buser1992,Obata1962}.

  The same viewpoint is naturally related to work on genuinely dynamical black hole mechanics and thermodynamics. Those approaches organise horizon evolution through first law, entropy, and geometric balance relations, while the present analysis develops the flux-side ray tracing constraint \cite{AshtekarKrishnan2004,HollandsWaldZhang2024,VisserYan2024,KongTianZhangZhao2025}.

When the centred antipodal odd component of the channel labelled peeling field is created during a finite retarded time interval and erased by its end, the time integrated channel defect is bounded below by a conformal forcing scale minus an integrated nonconformal remainder. The exact inequality concerns an extracted massless conformal channel. Relating the channel result to the complete Hawking luminosity at \(\mathscr I^+\) requires angular momentum barriers, greybody transmission, frequency mixing, mass and spin dependence, and interchannel coupling \cite{Page1976,Sanchez1978,MacGibbon1991}.

The local Schwarzian identity converts the retarded time derivative \(\partial_u\kappa_{\Sigma}\) into a conformal channel defect. The present result combines endpoint conditions with the largest centred antipodal mismatch to determine a profile-independent lower bound on the time-integrated channel norm. The exact statement applies to the massless conformal sector of the Hawking radiation problem. Four-dimensional angular barriers, greybody transmission, mass and spin dependence, frequency mixing, and interchannel coupling contribute through a separately defined remainder whose integrated norm requires independent control. The prescribed mismatch history provides the geometric input, while collapse, accretion, perturbation dynamics, and semiclassical back reaction determine its dynamical origin.

The central advance is the conversion of a global angular datum into a temporal forcing scale. The maximal centred antipodal mismatch records directional structure that can disappear from the slice average and at both temporal endpoints, while still fixing a profile-independent minimum integrated channel response. Balanced uniformisation supplies the intrinsic comparison map, and the Borsuk--Ulam theorem supplies an antipodal coincidence pair for continuous channel data. Together with the conformal flux identity, these ingredients yield a quantitative lower bound on the integrated conformal channel defect.

Throughout, we use units \(G=c=\hbar=k_{B}=1\).

\section{Ray tracing and the peeling field}\label{sec:peeling}

Let \(u\) be a retarded time coordinate on future null infinity \(\mathscr{I}^{+}\), and let \(U\) be an affine null parameter defined on the relevant ingoing side of the problem, for example on the past boundary or along an ingoing null family near the horizon. Outgoing null rays then determine a relation between these parameters. In spherical symmetry one writes simply
\begin{equation}
U=p(u).
\end{equation}
In a dynamical anisotropic setting the same relation acquires a direction label,
\begin{equation}
U=p(u,\hat n),
\label{eq:directional_p}
\end{equation}
where \(\hat n\in\mathbb{S}^{2}_{\infty}\) labels the outgoing asymptotic direction at \(\mathscr{I}^{+}\) \cite{Penrose1963,Sachs1962}.

Equation \eqref{eq:directional_p} is the basic kinematic input of the paper. It packages the accumulated redshift history seen by an outgoing channel. In stationary situations that history is encoded by an exponential relation near late retarded times. During a general transient episode, the function \(p(u,\hat n)\) may contain inseparable temporal and angular dependence, evolve across the full interval, and vary appreciably with \(\hat n\). Whenever a channelwise conformal reduction is valid, the renormalised flux is an exact functional of \(p\).

A basic regularity and causality requirement is monotonicity along each generator.
 \begin{equation}
\partial_{u}p(u,\hat n) > 0.
\label{eq:monotone_p}
\end{equation}
This ensures that the logarithm of \(\partial_{u}p\) is well defined and that the outgoing channel preserves the causal ordering inherited from the chosen affine parameter. For the flux law used later we also assume enough differentiability for the Schwarzian derivative to exist channelwise, which is why the standing assumptions require \(p(\,\cdot\,,\hat n)\in C^{3}(\mathcal I)\).

The peeling function \(\kappa(u,\hat n)\) is defined by
\begin{equation}
\kappa(u,\hat n) = -\partial_{u}\ln\!\big(\partial_{u}p(u,\hat n)\big),
\label{eq:kappa_def}
\end{equation}
or equivalently
\begin{equation}
\kappa(u,\hat n) = - \frac{\partial_{u}^{2}p(u,\hat n)}{\partial_{u}p(u,\hat n)}.
\label{eq:kappa_ratio_form}
\end{equation}
In stationary limits this reduces to the usual surface-gravity scale that controls the near-thermal spectrum \cite{Hawking1975,Unruh1976,Visser2003,BarceloLiberatiSonegoVisser2011}. Away from stationarity, it should be viewed more generally as the channelwise logarithmic rate of change of the null redshift map. The quantity \(\kappa(u,\hat n)\) is defined at \(\mathscr I^+\) through the ray tracing data. Horizon local prescriptions such as the Kodama--Hayward construction provide complementary geometric scales \cite{Hayward1994}. The extracted channel field introduced below additionally depends on the chosen horizon slice, its label transport prescription, the retarded time convention, and the asymptotic Bondi angular frame.

Several immediate properties are worth recording. The field \(\kappa\) is invariant under \(u\)-independent affine reparametrisations in each channel. If
 \begin{equation}
 \widetilde p(u,\hat n) = \alpha(\hat n)p(u,\hat n) + \beta(\hat n),
\qquad \alpha(\hat n) >0,
\end{equation}
where \(\alpha\) and \(\beta\) are independent of \(u\), then
 \begin{equation}
\partial_u\widetilde p = \alpha(\hat n) \,\partial_u p,
\end{equation}
and therefore
\begin{equation}
-\partial_u\ln(\partial_u\widetilde p) = -\partial_u\ln \!\bigl(\alpha (\hat n) \partial_u p \bigr)
 = -\partial_u\ln(\partial_u p) = \kappa.
\end{equation}
Thus \(\kappa\) depends only on the channelwise affine class of the ingoing parameter.

  The field \(\kappa\) also determines \(\partial_u p\) up to a positive multiplicative integration constant in each channel. Integrating \eqref{eq:kappa_def} from a reference time \(u_{\rm ref}\) gives
 \begin{equation}
\partial_{u}p(u,\hat n)
=
C_p (\hat n)\,
\exp\!\Bigg[
-\int_{u_{\rm ref}} ^{u}\kappa(\tilde u,\hat n)\,\mathrm{d}\tilde u
\Bigg],
\label{eq:pprime_from_kappa}
\end{equation}
with \(C_p(\hat n)>0\) fixed by affine normalisation. A second integration yields
\begin{equation}
p(u,\hat n)
=
p(u_{\rm ref},\hat n)
+
C_p (\hat n)\int_{u_{\rm ref}} ^{u}
\exp\!\Bigg[
-\int_{u_{\rm ref}} ^{s}\kappa(\tilde u,\hat n)\,\mathrm{d}\tilde u
\Bigg]\mathrm{d}s.
\label{eq:p_from_kappa}
\end{equation}
This makes clear that \(\kappa\) is the physically relevant differential data carried by the ray-tracing map, with the remaining freedom corresponding precisely to affine reparametrisation.

  The sign and time dependence of \(\kappa\) encode how the redshift history departs from stationarity. If \(\kappa\) is constant in a channel, then
\begin{equation}
\partial_u p \propto e^{-\kappa u},
\end{equation}
and one recovers the familiar late-time exponential relation associated with a stationary Hawking temperature. If \(\kappa\) varies in time, then the channel redshift is no longer governed by a single exponential scale, and the departure from instantaneous thermality in the conformal flux law is controlled exactly by \(\partial_u\kappa\). If \(\kappa\) varies with \(\hat n\), then different channels experience different redshift histories, and any attempt to summarise the dynamics by a single scalar temperature necessarily loses directional information.

For later use it is also convenient to note that the logarithmic form \eqref{eq:kappa_def} can be differentiated once more to give
\begin{equation}
\partial_u\kappa(u,\hat n)
=
-\partial_u^{2}\ln\!\big(\partial_u p(u,\hat n)\big),
\end{equation}
so that the conformal sector defect probes the retarded time variation of the channel redshift. The ray tracing map therefore provides the natural starting point, and the conformal stress tensor follows directly from \(p\), or equivalently from \(\kappa\).

The peeling field enters the analysis in two related roles. Sections~\ref{sec:peeling} and \ref{sec:conformal_flux} use the direction-resolved asymptotic quantity \(\kappa(u,\hat n)\). Section~\ref{sec:law} labels the physical channels by points \(x\in\Sigma\) on the fixed reference slice through the extraction map \(\Phi_u\). The composite ray tracing map is \(p_{\Sigma}(u,x)=p(u,\Phi_u(x))\), and its peeling field is \(\kappa_{\Sigma}(u,x)=-\partial_u\ln(\partial_u p_{\Sigma}(u,x))\), with \(\partial_u\) evaluated at fixed \(x\). This material derivative follows the extracted channel labelled by \(x\) and includes the evolution of \(\Phi_u\). The lower bound uses this channel-labelled peeling field on the fixed reference slice.

With this notation fixed, the exact conformal-sector flux identity may now be recalled.

\section{Conformal Hawking flux and exact Schwarzian identity}\label{sec:conformal_flux}

In a \(1+1\) conformal sector, assume that the incoming channel is in the conformal vacuum defined by the affine null coordinate \(U\), so that \(\langle T_{UU}\rangle=0\). The renormalised energy flux at \(\mathscr I^+\) is then an exact functional of the ray-tracing map \cite{FullingDavies1976,DaviesFulling1976,CarlitzWilley1987,ChristensenFulling1977,FabbriNavarro2005,DiFrancesco1997,GoodAndersonEvans2016}. For a monotone ray-tracing function \(U=p(u)\), one standard normalisation gives
\begin{equation}
F(u) = \langle T_{uu}\rangle = -\frac{c_{\mathrm{CFT}}}{24\pi}\,\{p(u),u\},
\label{eq:flux_schwarzian}
\end{equation}
where \(c_{\mathrm{CFT}}\) is the central charge, with \(c_{\mathrm{CFT}}=1\) for a single massless conformal scalar channel, and \(\{p,u\}\) is the Schwarzian derivative \cite{FullingDavies1976,DaviesFulling1976,FabbriNavarro2005,DiFrancesco1997}. The sign convention is fixed so that the monotone late-time stationary relation \[p(u)=U_H-C_H e^{-\kappa u}, \qquad C_H>0, \] for which \(p'(u)>0\), yields the positive thermal flux \(F=c_{\mathrm{CFT}}\kappa^{2}/(48\pi)\).

 For a general incoming state, the transformation law contains the additional state-dependent term

\begin{equation}
\langle T_{uu}(u)\rangle = \bigl[p'(u)\bigr]^2 \left. \langle T_{UU}(U)\rangle \right|_{U=p(u)} - \frac{c_{\mathrm{CFT}}}{24\pi}\{p(u),u\}.
\end{equation}

The conformal identity used below is the \(U\)-vacuum contribution.

 Here and below, primes denote derivatives with respect to \(u\).
\begin{equation}
\{p,u\} = \frac{p'''(u)}{p'(u)} - \frac{3}{2}\Big(\frac{p''(u)}{p'(u)}\Big)^{2}.
\end{equation}

Define
\begin{equation}
\kappa(u) = -\frac{p''(u)}{p'(u)}.
\end{equation}
Then
\begin{equation}
\frac{p''}{p'} = -\kappa,
\qquad
\frac{p'''}{p'} = \frac{d}{du}\Big(\frac{p''}{p'}\Big) + \Big(\frac{p''}{p'}\Big)^2 = -\kappa' + \kappa^{2}.
\end{equation}
Substituting into the Schwarzian gives
\begin{equation}
\{p,u\} = \big(-\kappa' + \kappa^{2}\big) - \frac{3}{2}\kappa^{2} = -\kappa' - \frac{1}{2}\kappa^{2},
\end{equation}
and therefore
\begin{equation}
F(u) = \frac{c_{\mathrm{CFT}}}{24\pi}\Big(\kappa'(u) + \frac{1}{2}\kappa(u)^{2}\Big).
\label{eq:flux_kappa}
\end{equation}

Define the quadratic reference term by retaining only the \(\kappa^{2}\) term \cite{Visser2003,BarceloLiberatiSonegoVisser2011,FabbriNavarro2005}.
 \begin{equation}
F_{\mathrm{th}}(u) = \frac{c_{\mathrm{CFT}}}{48\pi}\,\kappa(u)^{2}.
\label{eq:Fth_def}
\end{equation}
The exact conformal defect is then
\begin{equation}
F(u) - F_{\mathrm{th}}(u) = \frac{c_{\mathrm{CFT}}}{24\pi}\,\kappa'(u).
\label{eq:thermality_defect_identity}
\end{equation}
Equation~\eqref{eq:thermality_defect_identity} gives the nonadiabaticity identity. At retarded times for which \(\kappa(u)\neq0\), the quadratic reference contribution dominates in the regime \cite{Visser2003,BarceloLiberatiSonegoVisser2011}
\begin{equation}
\frac{|\kappa'(u)|}{\kappa(u)^2}\ll1.
\end{equation}
The ratio is undefined at a zero of \(\kappa\), where the derivative and quadratic terms must instead be compared directly. Its physical interpretation depends on the relation between the extracted conformal channel and the four-dimensional propagation problem.

  For a four-dimensional field decomposed into angular modes, the radial equation has the schematic form

\begin{equation}
\left[-\partial_t^2+\partial_{r_*}^2-V_{\ell s}(r) \right]\psi_{\ell m}=0.
\label{eq:four_dimensional_radial}
\end{equation}

For a scalar field of mass $\mu$ on a Schwarzschild background,

\begin{equation}
V_{\ell 0}^{(\mu)}(r) = \left(1-\frac{2M}{r}\right) \left[\frac{\ell(\ell+1)}{r^2} +\frac{2M}{r^3} +\mu^2 \right].
\label{eq:scalar_effective_potential}
\end{equation}

The Schwarzian identity gives the exact flux law for a massless conformal channel. This description arises directly in two-dimensional conformal models and locally through near-horizon reduction. In four dimensions, the effective potential controls the subsequent radial and angular propagation. A band-limited high-frequency regime provides a setting in which potential scattering can become weak, while a partial-wave description retains the propagation effects mode by mode.

The present theorem acts on extracted, direction-resolved conformal channels over a caustic-free interval. The first angularly structured contribution lies beyond the spherical \(\ell=0\) sector, for which \(\mathcal M_{\rm cent}=0\). Effectively massless packets, weak angular scattering, and controlled channel mixing provide the intended four-dimensional regime.

The local remainder \(F_{\mathrm{rem}}\) contains state-dependent contributions beyond the chosen conformal vacuum, greybody transmission and angular filtering \cite{Page1976,Sanchez1978,MacGibbon1991}, interchannel mixing, corrections to effective \(1+1\)-dimensional propagation, and residual extraction errors not already encoded in \(\Phi_u\). The integrated quantity \(\mathcal B_{\mathrm{rem}}\) measures their cumulative contribution over \(\mathcal I\). A four-dimensional application determines this quantity through an independent propagation calculation and compares it directly with the conformal forcing scale.

Let \(\mathcal{I}=[u_{0},u_{1}]\). Integrating the absolute defect gives
\begin{equation}
\int_{\mathcal{I}}\big|F(u)-F_{\mathrm{th}}(u)\big|\,\mathrm{d} u = \frac{c_{\mathrm{CFT}}}{24\pi}\int_{\mathcal{I}}\big|\kappa'(u)\big|\,\mathrm{d} u.
\label{eq:L1_gate}
\end{equation}

An illustrative perturbation about a stationary peeling background serves as a calibration of the nonadiabaticity identity. Let

\begin{equation}
\kappa(u) = \bar\kappa + \delta\kappa(u), \qquad \bar\kappa>0,\quad a_0>0,\quad \tau>0,
\label{eq:background_pulse}
 \end{equation}

  where

 \begin{equation}
 \delta \kappa (u) = \begin{cases}
a_0\sin^{2}\!\left[\dfrac{\pi(u-u_0)}{\tau} \right], &u_0\leq u\leq u_1,\\[6pt]
0,
&u\notin[u_0,u_1],
\end{cases}
  \qquad u_1=u_0+\tau.
\label{eq:compact_pulse}
 \end{equation}
 Its derivative is
\begin{equation}
\partial_u\kappa(u)
=
\partial_u\delta\kappa(u)
=
\begin{cases}
a_0\dfrac{\pi}{\tau} \sin\!\left[\dfrac{2\pi(u-u_0)}{\tau} \right], &u_0\leq u\leq u_1,\\[6pt] 0, &u\notin[u_0,u_1].
\end{cases}
\label{eq:compact_pulse_derivative}
\end{equation}
 The quadratic reference contribution is
 \begin{equation}
F_{\rm th} (u) = \frac{c_{\rm CFT}}{48\pi} \left[\bar\kappa+\delta\kappa(u) \right]^2,
\label{eq:compact_pulse_reference}
 \end{equation}
  and the exact conformal defect is
 \begin{equation}
F ^{\rm conf} (u)-F_{\rm th} (u) = \frac{c_{\rm CFT}}{24\pi} \partial_u\delta\kappa(u).
 \label{eq:compact_pulse_defect}
\end{equation}
  Therefore
 \begin{equation}
\int_{u_0}^{u_1} \left| F ^{\rm conf} (u)-F_{\rm th} (u) \right| \, \mathrm du = \frac{c_{\rm CFT}}{12\pi}a_0.
\label{eq:compact_pulse_integrated_defect}
\end{equation}
 The corresponding signed defect is
\begin{equation}
\int_{u_0}^{u_1} \left[F^{\rm conf}(u)-F_{\rm th}(u) \right] \,\mathrm du = \frac{c_{\rm CFT}}{24\pi} \left[\kappa(u_1)-\kappa(u_0) \right] = 0.
\label{eq:compact_pulse_signed_defect}
\end{equation}
Equation~\eqref{eq:compact_pulse_integrated_defect} gives the integrated absolute channel defect. Equal endpoint values of the peeling field give a zero signed integral in
\eqref{eq:compact_pulse_signed_defect}.

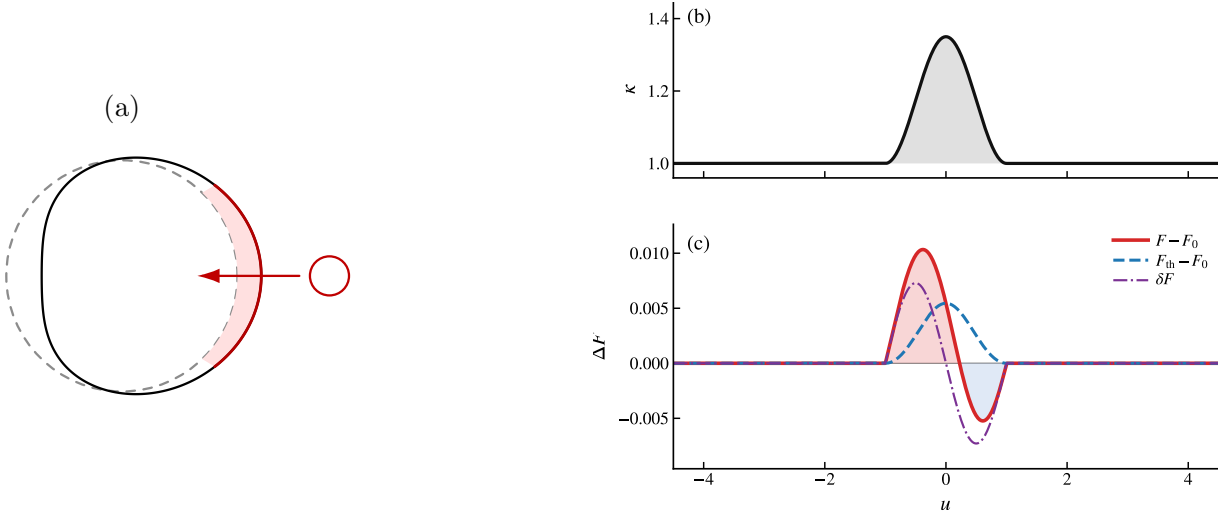
\begin{figure}
\centering

\begin{minipage}[c]{0.32\textwidth}
\centering
\vspace{0pt}

\resizebox{\linewidth}{!}{
\begin{tikzpicture}[
    >=Latex,
    line cap=round,
    line join=round
]

\def\xL{0.0}
\def\yC{0.0}
\def\R{1.42}

\useasboundingbox (-1.75,-1.80) rectangle (2.90,2.25);

\node at (\xL,2.05) {\small (a)};

\draw[dashed, line width=0.8pt, black!45]
    (\xL,\yC) circle (\R);

\fill[
    red!12,
    samples=180,
    smooth,
    variable=\t,
    domain=-46:46
]
  plot (
    {\xL + (\R*(1 + 0.24*cos(\t)
    - 0.050*cos(2*\t)
    + 0.016*cos(3*\t)))*cos(\t)},
    {\yC + (0.990*\R*(1 + 0.155*cos(\t)
    - 0.026*cos(2*\t)
    + 0.008*cos(3*\t)))*sin(\t)}
  )
  --
  plot[
    samples=180,
    smooth,
    variable=\t,
    domain=46:-46
  ]
  (
    {\xL + \R*cos(\t)},
    {\yC + \R*sin(\t)}
  )
  -- cycle;

\draw[
    thick,
    samples=220,
    smooth,
    variable=\t,
    domain=0:360
]
  plot (
    {\xL + (\R*(1 + 0.24*cos(\t)
    - 0.050*cos(2*\t)
    + 0.016*cos(3*\t)))*cos(\t)},
    {\yC + (0.990*\R*(1 + 0.155*cos(\t)
    - 0.026*cos(2*\t)
    + 0.008*cos(3*\t)))*sin(\t)}
  );

\draw[
    red!70!black,
    line width=1.0pt,
    samples=160,
    smooth,
    variable=\t,
    domain=-46:46
]
  plot (
    {\xL + (\R*(1 + 0.24*cos(\t)
    - 0.050*cos(2*\t)
    + 0.016*cos(3*\t)))*cos(\t)},
    {\yC + (0.990*\R*(1 + 0.155*cos(\t)
    - 0.026*cos(2*\t)
    + 0.008*cos(3*\t)))*sin(\t)}
  );

\draw[red!75!black, thick]
    (\xL+2.55,\yC) circle (0.24);

\draw[
    -{Latex[length=2.8mm,width=1.8mm]},
    thick,
    red!75!black
]
    (\xL+2.18,\yC) -- (\xL+0.92,\yC);

\end{tikzpicture}
}

\end{minipage}
\hfill
\begin{minipage}[c]{0.5\textwidth}
\centering
\vspace{0pt}

\includegraphics[
    width=\linewidth,
    trim=8 8 8 8,
    clip
]{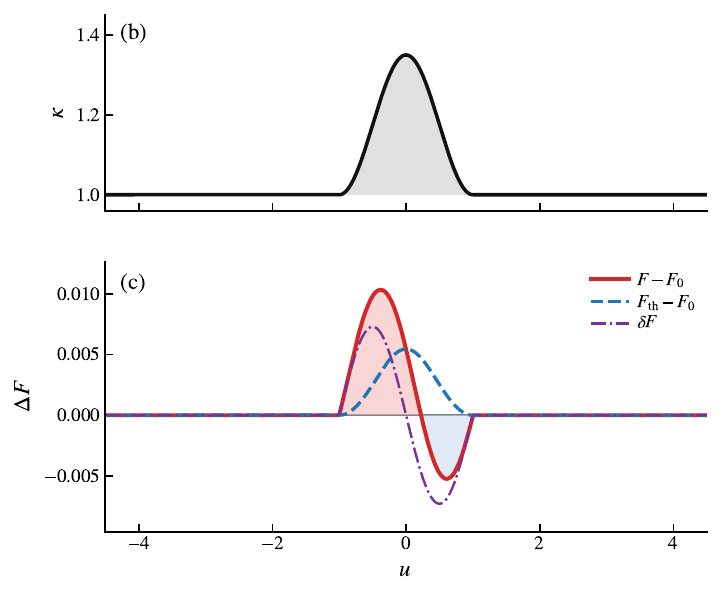}

\end{minipage}

 \caption{Panel~(a) shows the reference and perturbed horizon profiles.
Panel~(b) shows the full peeling profile
\(\kappa(u)=\bar\kappa+\delta\kappa(u)\), where only
\(\delta\kappa(u)\) has compact support.
Panel~(c) shows \(F^{\rm conf}(u)-F_0\), \(F_{\rm th}(u)-F_0\), and
\(\delta F(u)\equiv F^{\rm conf}(u)-F_{\rm th}(u)\) as red solid,
blue dashed, and purple dash-dotted curves, with
\(F_0=c_{\rm CFT}\bar\kappa^2/(48\pi)\).}
\label{fig:adiabaticity_gate}

 \end{figure}

Pulse-like flux profiles of this kind are also familiar in exact moving-mirror trajectories, where they provide a useful calibration language for transient nonadiabatic behaviour \cite{GoodAndersonEvans2013,GoodOng2015,GoodLinder2019}. Figure~\ref{fig:adiabaticity_gate} combines the transient angular structure with its conformal-channel response. Panel~(a) shows the asymmetric structure on the reference horizon slice. Panel~(b) shows the full peeling profile with a compact perturbation, and panel~(c) shows the baseline-subtracted exact flux, the quadratic reference contribution, and their defect. The same compact perturbation profile enters the round-sphere dipole calibration below.

\section{Intrinsic antipodal mismatch on a horizon slice}\label{sec:antipode}

Let \(\Sigma\) be a fixed smooth reference horizon cross-section with topology \(S^{2}\), intrinsic metric \(g_{ab}\), positive Riemannian area measure \(\mathrm{d}A\), and total area
\begin{equation}
\mathrm{Area}(\Sigma) = \int_{\Sigma} \mathrm{d} A.
\end{equation}
Uniformisation implies \((\Sigma,g)\) is conformal to the round unit sphere \(\mathbb{S}^{2}\) \cite{Jost2006,OsgoodPhillipsSarnak1988}. The sphere here provides the reference antipodal geometry against which antipodal-odd distortions can be measured. The construction fixes the selected horizon slice and its associated slicing throughout the analysis.

There exists a diffeomorphism \(h\) from \(\Sigma\) to \(\mathbb{S}^{2}\) and a smooth scalar field \(\rho\) such that
 \begin{equation}
h^{*}g^{\mathrm{rd}} = e ^{2\rho} g,
\end{equation}
where \(g^{\mathrm{rd}}\) denotes the round metric on \(\mathbb{S}^{2}\).
The residual freedom in \(h\) is Möbius. We fix it by a balanced gauge built from the pushed-forward normalised area measure, using the conformal barycenter idea in a form compatible with conformal naturality \cite{Hersch1970,DouadyEarle1986}.

Define the pushed-forward probability measure \(\nu_{h}\) on \(\mathbb{S}^{2}\) by
\begin{equation}
  \nu_{h}(B) = \frac{1}{\mathrm{Area}(\Sigma)} \int_{h^{-1}(B)}\mathrm{d} A.
\end{equation}
View \(\mathbb{S}^{2}\subset\mathbb{R}^{3}\) with coordinates \(y^{i}\).
Impose the barycenter condition
\begin{equation}
\int_{\mathbb{S}^{2}} y^{i}\,\mathrm{d} \nu_{h} = 0,
\end{equation}
and, when needed to break degeneracies, align the principal axes of
\begin{equation}
K^{ij} = \int_{\mathbb{S}^{2}} y^{i}y^{j}\,\mathrm{d} \nu_{h}.
\end{equation}
We assume a nondegenerate balanced representative. Residual round rotations leave the induced antipodal involution unchanged because the round antipode commutes with every round rotation. Symmetric degeneracies beyond this freedom require an explicit representative choice. The analysis follows one fixed nondegenerate balanced branch, with the associated distortion constants evaluated on that branch. Accordingly, the induced map below is well defined relative to the selected balanced branch; no branch-independent uniqueness is assumed in a degenerate case.

In this balanced gauge, define the induced antipodal map on \(\Sigma\) by pullback of the round antipode.
 \begin{equation}
A(x) = h^{-1}\!\big(-h(x)\big).
\label{eq:antipode_def}
\end{equation}
The map satisfies \(A(A(x))=x\). For every continuous map \(\mathbf v\) from \(\Sigma\) to \(\mathbb R^2\), the Borsuk--Ulam theorem supplies a point \(x\in\Sigma\) \cite{Matousek2003}. At that point,

\begin{equation}
\mathbf v(x)=\mathbf v(Ax).
\label{eq:antipodal_coincidence}
\end{equation}

For this pointwise statement, assume additionally that

\begin{equation}
\kappa_{\Sigma} \in C^{1}\!\bigl(\mathcal I;C^{0}(\Sigma)\bigr).
\label{eq:Borsuk_continuity}
\end{equation}

Thus both \(\kappa_{\Sigma}(u,\cdot)\) and
\(\partial_u\kappa_{\Sigma}(u,\cdot)\) are continuous scalar fields on \(\Sigma\) at every fixed \(u\). At fixed retarded time, set

\begin{equation}
\mathbf v_u(x)
=
\left(
\frac{\kappa_{\Sigma}(u,x)}{\kappa_{\rm ref}},
\frac{\partial_u\kappa_{\Sigma}(u,x)}{\kappa_{\rm ref}^2}
\right),
\qquad
\kappa_{\rm ref}>0.
\label{eq:channel_coincidence_map}
\end{equation}

Spatial continuity supplies an antipodal pair with equal peeling and equal peeling rate. The exact conformal flux agrees at the same pair. The theorem supplies at least one pair at each fixed retarded time, and the selected pair may vary with \(u\).

\begin{figure}
\centering
  \begin{tikzpicture}[>=stealth]

\draw[thick, fill=blue!5]
    (-1.8,0.6)
    .. controls (-1.2,1.8) and (1.4,1.6)..
    (2.2,0.2)
    .. controls (2.5,-1.2) and (0.7,-1.7)..
    (-1.4,-1.2)
    .. controls (-2.3,-0.5) and (-2.3,0.2)..
    cycle;

\node at (0,-1.7) {$(\Sigma,g)$};

\coordinate (x) at (-0.7,0.7);
\coordinate (Ax) at (1.0,-0.7);

\fill (x) circle (1.6pt);
\node[below right=1pt] at (x) {$x$};

\fill[red] (Ax) circle (1.6pt);
\node[left=3pt, red] at (Ax) {$A(x)$};

\coordinate (C) at (6.8,0);

\draw[thick] (C) circle (1.6cm);

\draw (5.2,0) arc (180:360:1.6 and 0.45);
\draw [dashed](8.4,0) arc (0:180:1.6 and 0.45);

\node at (6.8,-1.9) {$(\mathbb{S}^2, g^{\mathrm{rd}})$};

\coordinate (hx) at (6.0,0.8);
\coordinate (nhx) at (7.6,-0.8);

\draw[blue!85, densely dotted, semithick] (hx) -- (nhx);

\fill (hx) circle (1.6pt);
\node[right=2pt] at (hx) {$h(x)$};

\fill[red] (nhx) circle (1.6pt);
\node[left=2pt, red] at (nhx) {$-h(x)$};

\draw[->, thick, shorten >=4pt, shorten <=4pt]
    (1.8, 1.1) to[out=20, in=160]
    node[midway, above=2pt] {$h$}
    (5.4, 1.1);

\draw[->, thick, red, shorten >=4pt, shorten <=4pt]
    (5.4, -1.1) to[out=200, in=-20]
    node[midway, below=3pt] {$h^{-1}$}
    (1.8, -1.1);

\end{tikzpicture}

\caption{Balanced uniformisation construction of the induced antipodal map.}
\label{fig:canonical_antipode}
\end{figure}
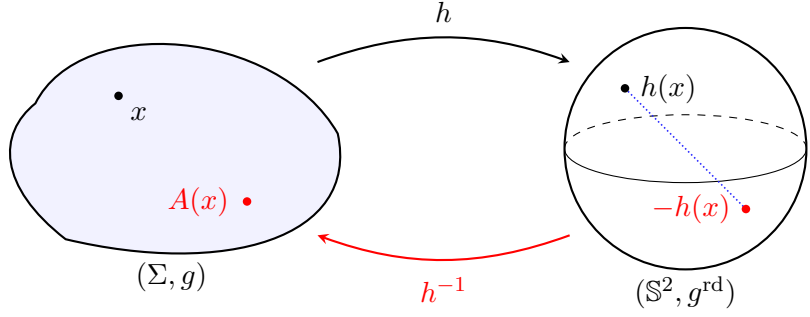

 The construction is summarised in Fig.~\ref{fig:canonical_antipode}. The balanced uniformisation map determines the induced antipodal map
 geometrically from the intrinsic two-geometry. Relative to the intrinsic two-geometry, \(A\) is the natural antipodal comparison map inherited from the round sphere after fixing the Möbius freedom.

Because the round antipode is an isometry of \((\mathbb S^2,g^{\mathrm{rd}})\) and \(h\) is conformal, the induced involution \(A=h^{-1}\circ(-\mathrm{id})\circ h\) is conformal with respect to \(g\). Indeed,

 \begin{equation}
  A^{*}g = e^{2\sigma}g, \qquad \sigma = \rho-\rho\circ A.
 \label{eq:A_conformal}
 \end{equation}

Here \(A^{*}\) denotes pullback by \(A\). Throughout, \(\mathrm dA\) denotes the positive Riemannian area measure, rather than an oriented area two-form. This distinction is relevant because the induced antipodal map is orientation reversing. Define the positive Jacobian \(J(x)\) by

\begin{equation}
A^{\ast}(\mathrm{d} A)\big|_{x} = J(x)\,\mathrm{d} A\big|_{x},
\label{eq:Jac_def}
\end{equation}

  so that

 \begin{equation}
J (x) = e^{2\sigma(x)}.
\end{equation}

 Define

\begin{equation}
J_{\min} = \min_{x\in\Sigma}J(x), \qquad J_{\max} = \max_{x\in\Sigma}J(x).
\end{equation}

 Because \(A\circ A=\mathrm{id}\),

\begin{equation}
J(A(x)) J(x)= 1,
 \label{eq:Jac_involution}
\end{equation}

 and consequently

\begin{equation}
J_{\max}=J_{\min}^{-1}.
\end{equation}

For a scalar \(f\) on \(\Sigma\), define its slice average
\begin{equation}
\savg{f} = \frac{1}{\mathrm{Area}(\Sigma)} \int_{\Sigma} f\,\mathrm{d} A,
\end{equation}
and similarly \(\savg{f\circ A}\). Here and below, composition is written as \((f\circ A)(x)=f(A(x))\).

Define the centred antipodal mismatch field
\begin{equation}
w_{f}(x) = f(x)-\savg{f} - f(A(x)) + \savg{f\circ A},
\end{equation}
which reduces to the usual antipodal-odd difference in the round-sphere isometric case. Define the associated mismatch functional
\begin{equation}
\mathcal{M}_{\mathrm{cent}}[f] = \int_{\Sigma} w_{f}(x)^{2}\,\mathrm{d} A.
\label{eq:Mcent_def}
\end{equation}

With the intrinsic involution and mismatch functional fixed, one can now quantify the angular cost of antipodal-odd structure.

\section{From antipodal mismatch to angular structure}\label{sec:spectral_cost}

\begin{figure}
\centering
\includegraphics[width=\textwidth]{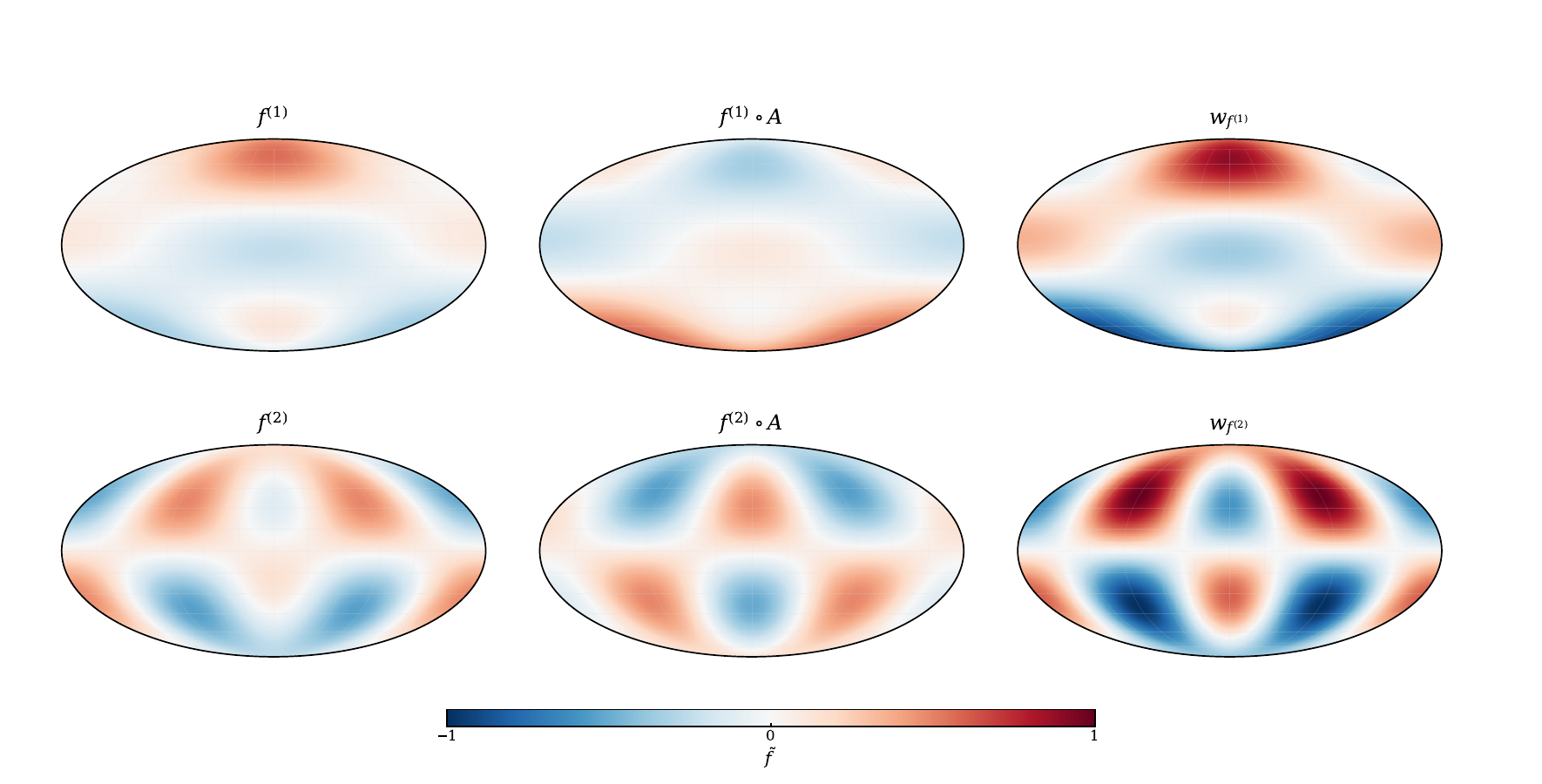}
\caption{Blue and red denote negative and positive normalised amplitude. The columns show each field, its antipodal image, and the centred mismatch.}
\label{fig:mismatch_decomposition}
\end{figure}

Before turning to the spectral estimate, it is useful to display the mismatch construction in the round-sphere limit for more than one representative morphology. Figure~\ref{fig:mismatch_decomposition} illustrates how the centred mismatch isolates the antipodal-odd part of a field while retaining its angular structure.

 For the spectral estimates in this section, let \(f\in H^{1}(\Sigma)\). Let \(\Delta=\nabla_{a}\nabla^{a}\) be the Laplacian on \((\Sigma,g)\).
Define the angular-gradient energy, i.e. the Dirichlet energy,

\begin{equation}
E[f] = \int_{\Sigma} g^{ab}\,\partial_{a}f\,\partial_{b}f\,\mathrm{d} A.
\label{eq:Dirichlet_energy}
\end{equation}

Let \(\lambda_{1}\) be the first nonzero eigenvalue of \(-\Delta\). For any \(q\in H^{1}(\Sigma)\) satisfying \(\int_{\Sigma}q\,\mathrm dA=0\),

 \begin{equation}
\int_{\Sigma} |\nabla q|^{2}\,\mathrm{d} A \ge \lambda_{1}\int_{\Sigma} q^{2}\,\mathrm{d} A,
\label{eq:poincare}
\end{equation}

a standard spectral-gap bound \cite{Chavel1984,Buser1992,Obata1962}. In applications where only a certified lower bound is available, write \(\lambda_{\rm cert}\le \lambda_{1}\) and replace \(\lambda_{1}\) by \(\lambda_{\rm cert}\) in the estimates below.

We relate \(w_{f}\) to \(f\) using distortion control, beginning with the effect of composition by the induced antipodal map on \(L^{2}\) norms and then turning to the corresponding effect on Dirichlet energy. For any \(\xi\in L^{2}(\Sigma)\), the change-of-variables formula associated with \(A^{*}(\mathrm{d}A)=J\,\mathrm{d}A\) gives

\begin{equation}
\int_{\Sigma} (\xi \circ A)^{2}\,\mathrm{d} A
=
\int_{\Sigma} \xi (y)^{2}\,J(y)\,\mathrm{d} A(y)
\le
J_{\max}\int_{\Sigma} \xi ^{2}\,\mathrm{d} A
=
J_{\min}^{-1}\int_{\Sigma} \xi ^{2}\,\mathrm{d} A,
\end{equation}

where the last equality uses the involutive identity \(J(A(x))J(x)=1\), which implies \(J_{\max}=J_{\min}^{-1}\). It follows that

\begin{equation}
\| \xi \circ A\|_{2} \le J_{\min}^{-1/2}\,\| \xi \|_{2}.
\label{eq:L2_comp_bound}
\end{equation}

For scalar fields in two dimensions, the Dirichlet energy is invariant under conformal changes of the metric. Since \(A\) is a conformal diffeomorphism of \((\Sigma,g)\),

 \begin{equation}
  E[f\circ A] = E[f].
\label{eq:energy_comp_bound}
\end{equation}

  Since derivatives annihilate constants, \(\nabla w_f=\nabla(f-f\circ A)\), and therefore

 \begin{equation}
  \begin{split}
\int_{\Sigma}|\nabla w_f|^2\,\mathrm dA \leq
2E[f]+2E[f\circ A] = 4E[f].
\end{split}
 \label{eq:w_energy_upper}
\end{equation}

  Combining \eqref{eq:w_energy_upper} with the spectral-gap bound \eqref{eq:poincare}, applied to the mean-zero field \(w_f\), gives

 \begin{equation}
E[f]
  \geq
\frac{\lambda_{\rm cert}}{4} \,
  \mathcal M_{\mathrm{cent}}[f].
\label{eq:odd_cost_bound}
\end{equation}

 The centred mismatch supplies global information that survives angular averaging. Combined with the endpoint conditions, its maximum value determines a profile-independent lower bound on the time-integrated channel norm. The spectral estimate therefore connects a nonlocal antipodal diagnostic to an intrinsic angular gradient scale.

 On a round sphere with areal radius \(R_{H}=\sqrt{\mathrm{Area}(\Sigma)/(4\pi)}\), \(\lambda_{1}=2/R_{H}^{2}\) and \(A\) is an isometry, giving the calibrated form \cite{Hersch1970,Obata1962}

\begin{equation}
E[f] \ge \frac{1}{2R_{H}^{2}}\,\mathcal{M}_{\mathrm{cent}}[f],
\label{eq:round_sharp}
\end{equation}

saturated by pure \(\ell=1\) antipodal-odd modes, where \(\ell\) is the spherical-harmonic degree in the round-sphere limit.

The calibration is explicit in the dipole sector.
On the round sphere of areal radius \(R_{H}\), let

\begin{equation}
f(\theta,\phi)=a\cos\theta.
\label{eq:dipole_mode}
\end{equation}

Then \(\savg{f}=0\) and \(f\circ A=-f\), so \(w_{f}=2f\).
Therefore

\begin{equation}
\mathcal{M}_{\mathrm{cent}}[f]
=
\frac{16\pi}{3}\,R_{H}^{2}a^{2},
\label{eq:dipole_mismatch}
\end{equation}

while

\begin{equation}
E[f]
=
\frac{8\pi}{3}\,a^{2}.
\label{eq:dipole_energy}
\end{equation}

Hence

\begin{equation}
E[f]=\frac{1}{2R_{H}^{2}}\,\mathcal{M}_{\mathrm{cent}}[f],
\label{eq:dipole_sharp}
\end{equation}

so the round-sphere constant is already sharp in the pure dipole sector. In the round-sphere dipole sector, the mismatch functional is exactly calibrated against the angular-gradient cost. Here the harmonic degree labels the scalar peeling field. The \(\ell=1\) mode supplies the sharp scalar calibration, while radiative spin two perturbations belong to the tensorial metric sector.

The calibration extends across the complete odd scalar sector. Taking a real spherical-harmonic basis, the harmonics obey

\begin{equation}
Y_{\ell m}(-\hat n) = (-1)^\ell Y_{\ell m}(\hat n).
\end{equation}

Every odd multipole therefore contributes fully to the centred antipodal mismatch. Combining this parity relation with the spherical Laplacian eigenvalue gives

\begin{equation}
Q_{\ell}
:=
\frac{E[Y_{\ell m}]}{\mathcal M_{\rm cent}[Y_{\ell m}]/(2R_H^2)}
=
\frac{\ell(\ell+1)}{2},
\qquad
\ell\ {\rm odd}.
\label{eq:odd_harmonic_ratio}
\end{equation}

The dipole saturates the round sphere estimate. Higher odd scalar multipoles carry a larger angular-gradient cost at fixed centred mismatch, with \(\ell=3\) providing the first higher-order scalar calibration. This statement concerns the scalar peeling field. Spin-two perturbations require a separate tensorial harmonic and parity analysis, which lies outside the scope of the present construction.

\section{Conditional bound for transient antipodal-odd structure}\label{sec:law}

Fix a time window \(\mathcal{I}=[u_{0},u_{1}]\) during which the outgoing sector can be modelled, channel by channel, by a conformal reduction with a direction-resolved peeling function \(\kappa(u,\hat n)\), together with the fixed reference horizon cross-section \((\Sigma,g)\) introduced above. Let \(\Phi_u\) be the channel-extraction map from \(\Sigma\) into \(\mathbb S^2_{\infty}\). The map assigns to each point $x\in\Sigma$ the outgoing asymptotic direction label traced along the chosen null congruence at retarded time $u$. We assume that \[\Phi \in C^{1}\!\bigl(\mathcal I\times \Sigma, \mathbb S^{2}_{\infty}\bigr), \] and that each fixed-\(u\) map \(\Phi_u:\Sigma\to\mathbb S^{2}_{\infty}\) is a \(C^{1}\) diffeomorphism. In practice \(\Phi_u\) is constructed by following outgoing null geodesics
from \(x\in\Sigma\) to \(\mathscr I^+\) and reading off the asymptotic angular coordinate. Operationally, the fixed reference slice supplies the channel labels and intrinsic geometry. The formalism does not select a unique quasilocal horizon or launch surface. The prescribed surface may be taken to be an event-horizon section, an apparent-horizon section, a stretched-horizon cut, or another fixed \(S^2\) launch surface, provided its intrinsic geometry
and label transport prescription are fixed.

A nearby timelike world tube carries corresponding cuts identified with \(\Sigma\), and outgoing rays launched from these cuts reach \(\mathscr I^+\). The world tube and its identification with \(\Sigma\) form part of the extraction prescription. Changing the prescribed surface or its label transport changes the extracted field \(\kappa_\Sigma\). On a caustic-free interval, the endpoint map remains single-valued, varies smoothly with the initial point, and preserves the direction-resolved channel labels. The distortion of \(\Phi_u\) belongs to the physical extraction prescription and is distinct from the intrinsic Jacobian \(J\) of the antipodal map \(A\).

To formulate the conformal law on the fixed reference slice, we label the outgoing channels by points \(x\in\Sigma\) and define the channel-labelled ray-tracing map

\begin{equation}
p_{\Sigma} (u,x)=p\bigl(u,\Phi_u(x)\bigr).
\label{eq:pS_def}
\end{equation}

Here \(x\) is the channel label, so differentiation with respect to \(u\) at fixed \(x\) is the material derivative along the extracted channel. We impose a channel-closure assumption. For every fixed \(x\in\Sigma\), the map \(u\mapsto p_{\Sigma}(u,x)\) is assumed to define the null-coordinate relation of an effective \(1+1\)-dimensional conformal channel. This closure is a modelling hypothesis and does not follow from a generic four-dimensional ray map when \(\Phi_u\) is time dependent. The Schwarzian flux law below is exact within the assumed extracted conformal channel. Angular transport induced by time-dependent relabelling, interchannel coupling, and nonconformal four-dimensional propagation that are not represented by this channelwise map are assigned to the remainder \(F_{\mathrm{rem}}\).

Throughout this section, we impose independently that the extracted channel map \(p_{\Sigma}(\cdot,x)\) lies in \(C^{3}(\mathcal I)\) and satisfies \(\partial_u p_{\Sigma}(u,x)>0\) for almost every \(x\in\Sigma\). The separate assumptions that \(p(u,\hat n)\) is \(C^{1}\) in its angular argument on the image of the extraction map and that \(\Phi\in C^{1}\!\bigl(\mathcal I\times\Sigma,\mathbb S^{2}_{\infty}\bigr)\) ensure that the first-derivative chain rule below is well defined; they are not used to infer the stated \(C^{3}\) regularity of \(p_{\Sigma}\). The family \(\Phi\) remains caustic free on \(\mathcal I\), with each \(\Phi_u\) a \(C^{1}\) diffeomorphism. The map \(u\mapsto\kappa_{\Sigma}(u,\cdot)\) is absolutely continuous as an \(L^{2}(\Sigma)\) -valued field. The induced antipodal map \(A\) is a \(C^{1}\) conformal involution with \(J_{\min}>0\), and \(F_{\mathrm{rem}}\in L^{1}\!\bigl(\mathcal I;L^{2}(\Sigma)\bigr)\).

Define the channel-labelled peeling field on the reference slice by
\begin{equation}
\kappa_{\Sigma} (u,x)
=
-\partial_u\ln\!\bigl(\partial_u p_{\Sigma} (u,x)\bigr)
=
- \frac{\partial_u^2 p_{\Sigma}(u,x)}{\partial_u p_{\Sigma}(u,x)}.
\label{eq:kappaS_material}
\end{equation}

  The field used below follows the transported channel label \(x\). Its derivative contains the time dependence of the extraction map through

\begin{equation}
\partial_u p_{\Sigma}(u,x) = (\partial_u p)\bigl(u,\Phi_u(x)\bigr) + \left. \mathrm d_{\hat n}p \right|_{(u,\Phi_u(x))} \!\left[\partial_u\Phi_u(x)\right].
\label{eq:pS_chainrule}
\end{equation}

The time dependence of \(\Phi_u\) therefore enters the composite ray-tracing map before the peeling field is formed. Defining \(\kappa_{\Sigma}\) directly from \(p_{\Sigma}\) gives the material field of the assumed extracted effective channel. Equation~\eqref{eq:pS_chainrule} displays only the first derivative; the \(C^3\) regularity needed by the Schwarzian law remains an independent hypothesis. Angular-transport and interchannel effects not captured by this channelwise representation are assigned to \(F_{\mathrm{rem}}\).

The balanced uniformisation map of Sec.~\ref{sec:antipode} defines the intrinsic antipodal involution \(A\), while \(\Phi_u\) supplies the physical channel extraction. The fields \(p_{\Sigma}\), \(\kappa_{\Sigma}\), and \(F_{\Sigma}\) are asymptotic channel quantities expressed using the fixed horizon labels.

  Under the channel-closure assumption above, define the exact flux of the extracted conformal channel on the fixed reference slice by

\begin{equation}
F_{\Sigma} ^{\mathrm{conf}}(u,x)
=
-\frac{c_{\mathrm{CFT}}}{24\pi}\,\{p_{\Sigma} (u,x),u\}.
\label{eq:Fs_conf_def}
\end{equation}

 We take \(F_{\Sigma}^{\mathrm{conf}}\) to denote the channelwise \(U\)-vacuum conformal contribution. State-dependent contributions beyond this conformal vacuum are included in the remainder defined below. Using \eqref{eq:kappaS_material}, the same algebra as in Sec.~\ref{sec:conformal_flux} gives

\begin{equation}
F_{\Sigma} ^{\mathrm{conf}}(u,x)
=
\frac{c_{\mathrm{CFT}}}{24\pi}
\Bigl(
\partial_u \kappa_{\Sigma} (u,x)
+
\frac{1}{2}\kappa_{\Sigma} (u,x)^2
\Bigr).
\label{eq:Fs_conf_kappa}
\end{equation}

Define the quadratic term in the conformal flux law by

\begin{equation}
F_{\mathrm{th}}(u,x)=\frac{c_{\mathrm{CFT}}}{48\pi}\,\kappa_{\Sigma} (u,x)^2,
\label{eq:Fth_slice}
\end{equation}

so that the exact conformal-channel defect is

\begin{equation}
\delta F_{\Sigma} ^{\mathrm{conf}}(u,x)
 =
F_{\Sigma} ^{\mathrm{conf}}(u,x)-F_{\mathrm{th}}(u,x)
=
\frac{c_{\mathrm{CFT}}}{24\pi}\,\partial_u \kappa_{\Sigma} (u,x).
\label{eq:defect_conf_material}
\end{equation}

Let the physical channel-labelled flux on the reference slice be \(F_{\Sigma}(u,x)\), and define

\begin{equation}
\delta F_{\Sigma} (u,x)=F_{\Sigma} (u,x)-F_{\mathrm{th}}(u,x).
\end{equation}

We then write

\begin{equation}
\delta F_{\Sigma} (u,x)
=
\frac{c_{\mathrm{CFT}}}{24\pi}\,\partial_u \kappa_{\Sigma} (u,x)
+
  F_{\mathrm{rem}} (u,x),
\label{eq:defect_with_budget}
\end{equation}

where \(F_{\mathrm{rem}}\) collects state-dependent contributions beyond the chosen conformal vacuum, greybody transmission and angular filtering \cite{Page1976,Sanchez1978,MacGibbon1991}, interchannel mode mixing, time-dependent angular-transport effects not represented by the effective channel map, corrections to effective \(1+1\) -dimensional propagation, and residual extraction errors not already contained in the composite map \(p_{\Sigma}=p\circ\Phi_u\). Define the integrated remainder budget by

\begin{equation}
\mathcal B_{\mathrm{rem}}
=
\int_{\mathcal I}
\left\|F_{\mathrm{rem}}(u,\cdot)\right\|_2
\,\mathrm du.
\label{eq:integrated_remainder_budget}
\end{equation}

An independent four-dimensional propagation calculation supplies \(\mathcal B_{\mathrm{rem}}\). Quantitative applications compare this budget with the conformal forcing scale. For applications, the budget can be reported through separate state-dependent, greybody, angular-transport, and extraction contributions before they are combined into \(\mathcal B_{\mathrm{rem}}\). This bookkeeping identifies the physical origin of the propagation correction and allows each contribution to be estimated independently.

A useful high-frequency regime is indicated schematically by

\begin{equation}
\omega_cR_H\gg1,
\qquad
\omega_c^2
\gg
\sup_{\ell\in\mathcal L}\sup_{r_*}
\left|V_{\ell s}(r_*)\right|,
\qquad
\chi\ll1,
\qquad
\frac{\mu}{\omega_c}\ll1.
\label{eq:schematic_regime}
\end{equation}

  Here \(\omega_c\) denotes the characteristic packet frequency, \(\mathcal L\) is the retained partial-wave set, \(\chi\) measures the nonspherical distortion, and \(\mu\) is the field mass. The potential-height condition and \(\omega_cR_H\gg1\) are regime indicators for weak scattering in the retained high-frequency modes; they do not by themselves bound reflection, angular mixing, or the full propagation correction. They also select a high-frequency subset of the Hawking spectrum. The quantity \(\mathcal B_{\mathrm{rem}}\) must therefore still be supplied by an independent four-dimensional propagation calculation.

We now specialise to a round reference cross-section of areal radius \(R_H\) and a dipole-dominated extracted channel field. Let

\begin{equation}
\kappa_{\Sigma} (u,\theta,\phi)
=
 \bar\kappa+ a(u)\cos\theta,
\qquad
 0< a_0<\bar\kappa.
 \label{eq:dipole_pulse_model}
\end{equation}

  The pulse amplitude is

\begin{equation}
a(u)
=
\begin{cases}
a_0\sin^{2}\!\left[
\dfrac{\pi(u-u_0)}{\tau}
\right],
&
u_0\leq u\leq u_1,
\\[6pt]
0,
&
u\notin[u_0,u_1],
\end{cases}
\qquad
u_1=u_0+\tau.
\label{eq:dipole_amplitude}
\end{equation}

The full peeling field satisfies

\begin{equation}
\kappa_{\Sigma}(u,\theta,\phi)
\geq
\bar\kappa-a_0
>
0.
\label{eq:positive_dipole_peeling}
\end{equation}

 Define the extracted ray-tracing map by

\begin{equation}
p_{\Sigma} (u,\theta,\phi)
=
\int_{u_{\rm ref}} ^{u}
\exp\!\Bigg[
-\int_{u_{\rm ref}} ^{s}\kappa_{\Sigma} (\tilde u,\theta,\phi)\,\mathrm{d}\tilde u
\Bigg]\mathrm{d}s,
\label{eq:model_pS_explicit}
\end{equation}

so that $\partial_u p_{\Sigma}>0$ and $p_{\Sigma}(\cdot,\theta,\phi)\in C^3(\mathcal I)$.

For this model, the exact conformal-channel defect is

\begin{equation}
\delta F_{\Sigma} ^{\mathrm{conf}}(u,\theta,\phi)
=
\frac{c_{\mathrm{CFT}}}{24\pi}\,\partial_u\kappa_{\Sigma} (u,\theta,\phi)
=
\frac{c_{\mathrm{CFT}}}{24\pi}\,a'(u)\cos\theta.
\label{eq:defect_conf_dipole_model}
\end{equation}

This construction provides an exact conformal calibration of the bound. The prescribed pulse fixes its normalisation and demonstrates saturation in the round sphere dipole sector.

On the round sphere,

\begin{equation}
\|\cos\theta\|_{2}
= \Big(\int_{\Sigma} \cos^{2}\theta\,\mathrm{d}A\Big)^{1/2} = \Big(\frac{4\pi R_H^{2}}{3}\Big)^{1/2},
\end{equation}

hence

\begin{equation}
\|\delta F_{\Sigma} ^{\mathrm{conf}}(u,\cdot)\|_{2}
=
\frac{c_{\mathrm{CFT}}}{24\pi}
|a'(u)|
\Big(\frac{4\pi R_H^{2}}{3}\Big)^{1/2}.
\label{eq:defect_conf_norm_model}
\end{equation}

Since $a(u)$ rises from $0$ to $a_0$ and returns to $0$ on $\mathcal I$,

\begin{equation}
\int_{\mathcal I}|a'(u)|\,\mathrm{d}u = 2a_0,
\end{equation}

and therefore

\begin{equation}
\int_{\mathcal I}\|\delta F_{\Sigma} ^{\mathrm{conf}}(u,\cdot)\|_{2}\,\mathrm{d}u
=
\frac{c_{\mathrm{CFT}}}{12\pi}
\Big(\frac{4\pi R_H^{2}}{3}\Big)^{1/2} a_0.
\label{eq:integrated_conf_defect_model}
\end{equation}

  The round-sphere antipode gives

 \begin{equation}
 \savg{\kappa_{\Sigma}(u,\cdot)}
=
\bar\kappa,
\qquad
 w_{\kappa}(u, \theta, \phi)
  =
2a (u) \cos\theta.
\label{eq:dipole_centred_field}
 \end{equation}

  Consequently,

 \begin{equation}
  \mathcal M_{\mathrm{cent}}
[\kappa_{\Sigma} (u, \cdot)]
 =
 \frac{16\pi}{3}R_H^2a (u) ^2,
 \qquad
\mathcal M_{\max}
=
\frac{16\pi}{3}R_H^2a_0^2.
\label{eq:dipole_mismatch_history}
 \end{equation}

  For \(J_{\min}=1\), the conformal forcing scale becomes

 \begin{equation}
  \mathcal S_{\mathrm{conf}}
=
\frac{c_{\mathrm{CFT}}}{24\pi}
 \sqrt{\mathcal M_{\max}}
=
\frac{c_{\mathrm{CFT}}}{12\pi}
  \left(
\frac{4\pi R_H^2}{3}
\right) ^{1/2}
  a_0.
\label{eq:Sconf_model}
\end{equation}

Comparison with Eq.~\eqref{eq:integrated_conf_defect_model} gives

 \begin{equation}
\int_{\mathcal I}
\|\delta F_{\Sigma} ^{\mathrm{conf}}(u,\cdot)\|_2
 \, \mathrm du
 =
\mathcal S_{\mathrm{conf}}.
\label{eq:saturation_model}
\end{equation}

  The signed temporal integral vanishes in every channel, and the angular integral vanishes at each retarded time.

 For the pulse calibration, define the normalised cumulative absolute response by

\begin{equation}
\mathcal C(u)
=
\frac{
\displaystyle\int_{u_0}^{u}|a'(v)|\,\mathrm dv
}{
\displaystyle\int_{u_0}^{u_1}|a'(v)|\,\mathrm dv
},
\qquad
u\in\mathcal I.
\label{eq:cumulative_response}
\end{equation}

\begin{figure}
\centering
\includegraphics[width=\textwidth]{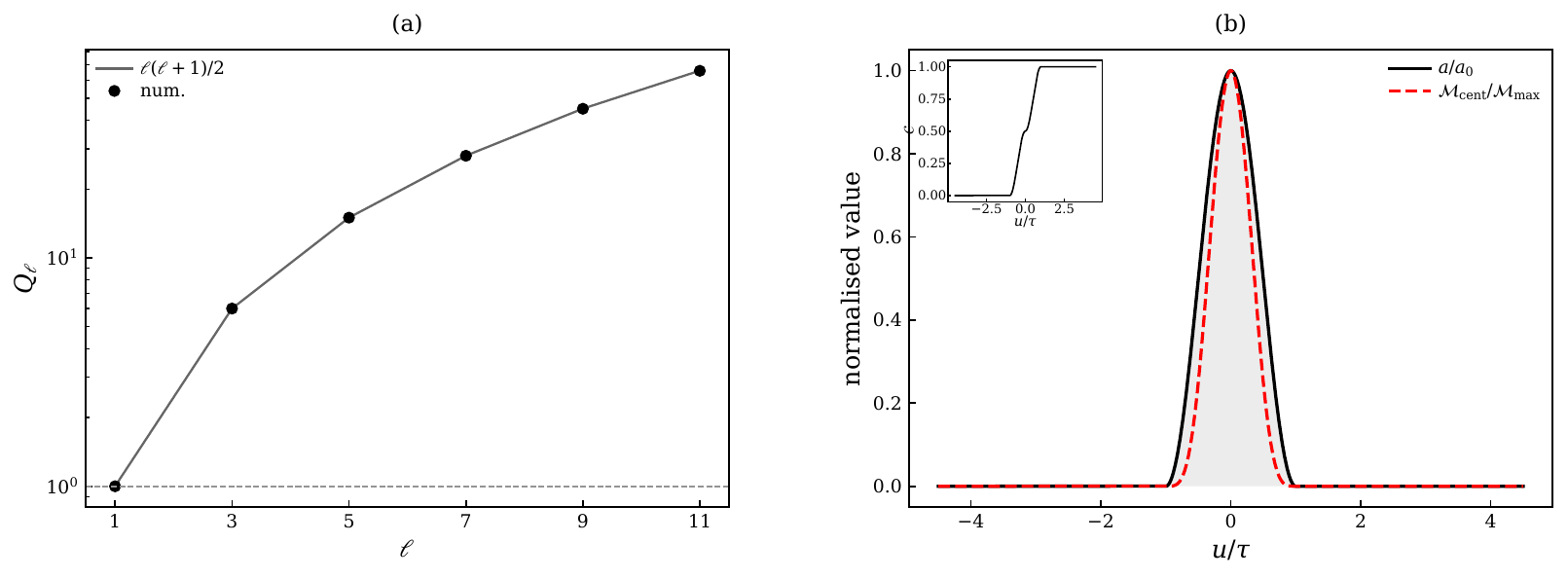}
\caption{Panel~(a) compares the exact and numerical values of \(Q_{\ell}\). Panel~(b) shows \(a/a_0\) and \(\mathcal M_{\rm cent}/\mathcal M_{\max}\) as functions of \(u/\tau\). The inset shows the normalised cumulative absolute response \(\mathcal C(u)\).}
\label{fig:quantitative_calibration}
\end{figure}

 Figure~\ref{fig:quantitative_calibration} shows the odd-harmonic calibration, the transient dipole history, and \(\mathcal C(u)\).

We collect the assumptions used in the bound. Let \((\Sigma,g)\) be a fixed smooth reference cross-section with induced conformal antipodal involution \(A\), and take \(c_{\mathrm{CFT}}>0\). Let \(p_{\Sigma}(u,x)\) be a caustic-free extracted channel map on \(\mathcal I\times\Sigma\), with \(p_{\Sigma}(\cdot,x)\in C^{3}(\mathcal I)\) and \(\partial_u p_{\Sigma}>0\) for almost every \(x\). For the integrated lower bound, assume that \(u\mapsto\kappa_{\Sigma}(u,\cdot)\) is absolutely continuous as an \(L^{2}(\Sigma)\) -valued field and that \(F_{\mathrm{rem}}\in L^{1}(\mathcal I;L^{2}(\Sigma))\) satisfies Eq.~\eqref{eq:defect_with_budget}. The pointwise coincidence statement additionally uses Eq.~\eqref{eq:Borsuk_continuity}; that stronger spatial-continuity hypothesis is not required for the integrated \(L^{2}\) bound. These assumptions yield the lower bound in Eq.~\eqref{eq:law_main} and the robust form in Eq.~\eqref{eq:law_robust}.

The minimal transient condition is

 \begin{equation}
\mathcal{M}_{\mathrm{cent}}[\kappa_{\Sigma}(u_{0},\cdot)] = 0,\qquad
\mathcal{M}_{\mathrm{cent}}[\kappa_{\Sigma}(u_{1},\cdot)] = 0.
\label{eq:created_erased}
\end{equation}

Define \(L^{2}(\Sigma)\) norms at fixed \(u\) by

\begin{equation}
\|X(u,\cdot)\|_{2} = \Big(\int_{\Sigma} X(u,x)^{2}\,\mathrm{d} A\Big)^{1/2}.
\end{equation}

A standard total-variation estimate, applied to an absolutely continuous \(L^{2}(\Sigma)\) -valued field \(q(u,\cdot)\), implies that if \(q(u_{0})=0=q(u_{1})\), then

\begin{equation}
\int_{\mathcal{I}}\|\partial_{u}q(u,\cdot)\|_{2}\,\mathrm{d} u \ge 2\sup_{u\in \mathcal{I}}\|q(u,\cdot)\|_{2},
\label{eq:total_variation_bound}
\end{equation}

which follows from the fundamental theorem of calculus and the triangle inequality \cite{EvansGariepy2015}.
Apply this to \(q=w_{\kappa}\) and define

\begin{equation}
\mathcal{M}_{\max} = \sup_{u\in \mathcal{I}}\,\mathcal{M}_{\mathrm{cent}}[\kappa_{\Sigma}(u,\cdot)].
\end{equation}

Then \(\sup_{u}\|w_{\kappa}(u,\cdot)\|_{2} = \sqrt{\mathcal{M}_{\max}}\), and

\begin{equation}
\int_{\mathcal{I}}\|\partial_{u}w_{\kappa}(u,\cdot)\|_{2}\,\mathrm{d} u \ge 2\sqrt{\mathcal{M}_{\max}}.
\label{eq:w_total_variation}
\end{equation}

Relate \(\partial_{u}w_{\kappa}\) to \(\partial_{u}\kappa_{\Sigma}\).
Since \(w_{\kappa}\) is a centred difference of \(\kappa_{\Sigma}\) and \(\kappa_{\Sigma}\circ A\), and orthogonal projection onto the mean-zero subspace is contractive in the \(L^{2}\) norm,

\begin{equation}
\|\partial_{u}w_{\kappa}(u,\cdot)\|_{2}
\le
\|\partial_{u}\kappa_{\Sigma} (u,\cdot) - \partial_{u}\kappa_{\Sigma} (u,\cdot)\circ A\|_{2}
\le
\Big(1+J_{\min}^{-1/2}\Big)\,\|\partial_{u}\kappa_{\Sigma} (u,\cdot)\|_{2},
\label{eq:wprime_kprime}
\end{equation}

using \eqref{eq:L2_comp_bound}.
Combine \eqref{eq:w_total_variation} and \eqref{eq:wprime_kprime} to obtain

\begin{equation}
\int_{\mathcal{I}}\|\partial_{u}\kappa_{\Sigma} (u,\cdot)\|_{2}\,\mathrm{d} u
\ge
\frac{2}{1+J_{\min}^{-1/2}}\,\sqrt{\mathcal{M}_{\max}}.
\label{eq:kprime_lower}
\end{equation}

  Integrating the defect model in Eq.~\eqref{eq:defect_with_budget} and applying the triangle inequality gives

 \begin{equation}
\|\delta F_{\Sigma} (u,\cdot)\|_{2}
\ge
\frac{c_{\mathrm{CFT}}}{24\pi}
 \|\partial_{u}\kappa_{\Sigma} (u,\cdot)\|_{2}
-
\| F_{\mathrm{rem}} (u,\cdot)\|_{2}.
\end{equation}

  Define the integrated channel defect by

\begin{equation}
\mathcal D_{\mathcal I}
=
\int_{\mathcal I}
\|\delta F_{\Sigma}(u,\cdot)\|_2
\,\mathrm du.
\label{eq:integrated_channel_defect}
\end{equation}

Integration over \(\mathcal I\), together with Eq.~\eqref{eq:kprime_lower}, yields

 \begin{equation}
  \boxed{
\mathcal D_{\mathcal I}
\ge
\frac{c_{\mathrm{CFT}}}{24\pi}
\frac{2}{1+J_{\min}^{-1/2}}
\sqrt{\mathcal M_{\max}}
-
\mathcal B_{\mathrm{rem}}
}
 \label{eq:law_main}
\end{equation}

The bound applies to the extracted conformal-channel description with a controlled integrated remainder budget.

Condition \eqref{eq:created_erased} describes exact endpoint restoration. Approximate restoration is represented by

 \begin{equation}
\mathcal{M}_{\mathrm{cent}}[\kappa_{\Sigma}(u_0,\cdot)] \le \epsilon_{\rm end}^{2},
\qquad
\mathcal{M}_{\mathrm{cent}}[\kappa_{\Sigma}(u_1,\cdot)] \le \epsilon_{\rm end}^{2}.
 \label{eq:approx_transient}
\end{equation}

  The same steps yield the robust lower bound

\begin{equation}
  \mathcal D_{\mathcal I}
 \ge
\frac{c_{\mathrm{CFT}}}{24\pi}\,
\frac{2}{1+J_{\min}^{-1/2}}\,
\bigl(\sqrt{\mathcal{M}_{\max}}- \epsilon_{\rm end} \bigr)
- \mathcal B_{\mathrm{rem}}.
 \label{eq:law_robust}
\end{equation}

  The right-hand side is positive whenever
$\sqrt{\mathcal{M}_{\max}}>\epsilon_{\rm end}
+ \tfrac{24\pi}{c_{\mathrm{CFT}}}(1+J_{\min}^{-1/2})\mathcal B_{\mathrm{rem}}/2$.
Setting \(\epsilon_{\rm end}=0\) recovers Eq.~\eqref{eq:law_main}.

  Transient creation and erasure of antipodal-odd structure generate the conformal forcing scale

\begin{equation}
  \mathcal S_{\mathrm{conf}} = \frac{c_{\mathrm{CFT}}}{24\pi}\,
\frac{2}{1+J_{\min}^{-1/2}}\,
\sqrt{\mathcal{M}_{\max}}.
\end{equation}

Then \eqref{eq:law_main} becomes

\begin{equation}
  \mathcal D_{\mathcal I} \ge \mathcal S_{\mathrm{conf}}- \mathcal B_{\mathrm{rem}}.
\end{equation}

The natural control parameter is therefore

\begin{equation}
\eta_{\mathrm{rem}} = \frac{\mathcal B_{\mathrm{rem}}}{\mathcal S_{\mathrm{conf}}},
\qquad
\mathcal S_{\mathrm{conf}}>0.
 \end{equation}

  When \(\mathcal S_{\mathrm{conf}}>0\), the condition \(\eta_{\mathrm{rem}}<1\) gives a positive lower bound. If \(\mathcal M_{\max}=0\), then \(\mathcal S_{\mathrm{conf}}=0\) and this ratio is not defined.

In applications, one fixes a horizon cross-section \((\Sigma,g)\), constructs the balanced-uniformisation antipodal map \(A\), traces a caustic-free family of outgoing null generators to \(\mathscr I^{+}\), and thereby obtains \(\Phi_u\), the extracted channel map \(p_{\Sigma}(u,\cdot)\), and the channel-labelled peeling field \(\kappa_{\Sigma}(u,\cdot)\). The mismatch history determines \(\mathcal M_{\max}\), while the intrinsic antipodal map supplies \(J_{\min}\). Independently, the spectral lower bound \(\lambda_{\rm cert}\) controls the conversion from centred mismatch to angular-gradient energy. The practical content of the inequality is the comparison between the conformal forcing scale \(\mathcal S_{\mathrm{conf}}\) and the independently estimated remainder budget \(\mathcal B_{\mathrm{rem}}\).

On a round cross-section with areal radius \(R_{H}\), the induced antipodal map is an isometry and the spectral floor is \(\lambda_{1}=2/R_{H}^{2}\) \cite{Hersch1970,Obata1962}. In that limit, pure dipole fields in the \(\ell=1\) antipodal-odd sector saturate the mismatch-to-gradient conversion in Eq.~\eqref{eq:round_sharp}.

When the antipodal-odd component of \(\kappa_{\Sigma}(u,\cdot)\) is dipole dominated, the mismatch-to-gradient step is sharp, so that in this regime the lower bound \eqref{eq:law_main} is controlled directly by the total variation of the dipole amplitude.

 Let

\begin{equation}
\tau=u_1-u_0
\end{equation}

denote the duration of the general transient interval \(\mathcal I=[u_0,u_1]\). For any integrable scalar function \(X(u)\), define the interval average

\begin{equation}
\langle X\rangle_{\mathcal I}
:=
\frac{1}{\tau}\int_{\mathcal I}X(u)\,\mathrm du.
\label{eq:interval_average}
\end{equation}

Define the area-normalised antipodal peeling scale by

\begin{equation}
\kappa_{\rm ap}
:=
\left(
\frac{\mathcal M_{\max}}{\operatorname{Area}(\Sigma)}
\right)^{1/2}.
\label{eq:antipodal_peeling_scale}
\end{equation}

Dividing the integrated bound by the episode duration and the square root of the horizon area gives

\begin{equation}
\boxed{
\frac{
\left\langle\|\delta F_{\Sigma}(u,\cdot)\|_2\right\rangle_{\mathcal I}
+
\left\langle\|F_{\mathrm{rem}}(u,\cdot)\|_2\right\rangle_{\mathcal I}
}{
\sqrt{\operatorname{Area}(\Sigma)}
}
\geq
\frac{c_{\rm CFT}}{12\pi}
\frac{\kappa_{\rm ap}}{\tau\left(1+J_{\min}^{-1/2}\right)}
}
\label{eq:time_averaged_channel_bound}
\end{equation}

For a round horizon slice, the scalar dipole calibration reaches equality in the conformal limit. The characteristic transient channel scale is therefore set by the area-normalised antipodal peeling amplitude divided by the episode duration.

 \section{Discussion}\label{sec:discussion_conclusions}

We have derived a lower bound on the time-integrated defect from the quadratic term in the conformal flux law in terms of transient antipodal-odd structure of the channel-labelled peeling field on a fixed reference cross-section. The maximal centred antipodal mismatch sets the forcing scale, while the intrinsic Jacobian distortion of the induced antipode and the integrated propagation budget determine its geometric and physical degradation.

The central geometric ingredient is the antipodal involution induced by balanced uniformisation, together with the associated centred mismatch functional. This pair detects odd angular structure directly on a distorted horizon cross-section and retains information beyond slice averages and coordinate-based identifications. Within the extracted conformal description, the mismatch becomes a quantitative forcing term for the integrated conformal-channel defect.

  Creation and erasure of nonzero antipodal odd peeling structure impose a positive forcing scale on the time-integrated spatial norm of the conformal channel defect. The conformal forcing scale measures the contribution fixed by the transient geometry, while \(\mathcal B_{\mathrm{rem}}\) records the cumulative propagation correction.

The controlled quantity is the unsigned time-integrated spatial \(L^{2}\) norm. The compact pulse exhibits a finite absolute response together with a vanishing signed defect, resolving transient channel activity and signed energy balance as distinct diagnostics.

The round-sphere dipole sector calibrates the general estimate. The geometric and temporal steps both reach equality, fixing the normalisation of the conformal forcing scale. Higher odd scalar harmonics carry a larger angular-gradient cost at fixed centred mismatch.

The result gives a structural relation between the extracted channelwise ray tracing map, its peeling field, and the transient horizon geometry. These data retain directional information that disappears under angular averaging.

  Four-dimensional applications combine a caustic-free extraction map with an independent estimate of the integrated remainder. Greybody propagation, frequency mixing, angular coupling, mass effects, and spin dependence connect the extracted channel response to the four-dimensional outgoing Hawking flux.

  Related studies of Hawking radiation use tunnelling methods and examine particle production, scattering, and thermodynamics in black holes with matter environments, regular cores, or noncommutative corrections \cite{VanzoAcquavivaDiCriscienzo2011,HeidariAraujoBarros2026,AraujoHeidariLobo2025,AraujoFilho2025LorentzianNC,Heidari2026NCCharged,MalufNeves2018}. Complementary phenomenological work uses black hole evaporation to constrain additional particle degrees of freedom and primordial black hole populations, and examines tunnelling to white hole configurations \cite{EwasiukProfumo2025DarkDOF,KorwarProfumo2023,EwasiukProfumo2026WhiteHole}. The present construction complements these approaches by constraining the ray-tracing response generated by transient angular structure.

The time-averaged form in Eq.~\eqref{eq:time_averaged_channel_bound} expresses the transient conformal channel scale through the area-normalised antipodal peeling amplitude and the episode duration. The round sphere dipole calibration reaches equality in the conformal limit, fixing the normalisation of this scaling.

  Perturbative collapse, asymmetric accretion, and numerical black-hole mergers provide possible settings in which \(\Phi_u\), \(\kappa_{\Sigma}\), \(\mathcal M_{\max}\), and \(\mathcal B_{\mathrm{rem}}\) can be evaluated. Greybody propagation and extensions across spin and frequency sectors then connect the geometric forcing scale to transmitted four-dimensional flux channels. The present theorem therefore supplies a reusable diagnostic linking
transient horizon geometry to the minimum integrated response of an extracted massless conformal channel associated with Hawking radiation.

\section*{Acknowledgements}
The authors thank the anonymous referees for their careful and constructive comments, which improved the clarity, scope, and physical interpretation of the manuscript.

\section*{Conflict of interest}
The authors declare that they have no conflict of interest.

\section*{Funding}
Self-funded research.

\section*{Author contributions}
Both authors contributed to conceptualization, derivations, and writing.

\section*{Data availability}
No external datasets were used. The numerical values shown in the figures were generated directly from the equations in the manuscript.

\bibliographystyle{iopart-num}
\bibliography{references}

\end{document}